\documentstyle[12pt]{article}

\newcommand{\sect}[1]{\setcounter{equation}{0}\section{#1}\indent}

\newcommand{\EQ}{\begin{equation}}
\newcommand{\EN}{\end{equation}}
\newcommand{\bea}{\begin{eqnarray}}
\newcommand{\ena}{\end{eqnarray}}
\newcommand{\vs}[1]{\vspace{#1 mm}}

\renewcommand{\a}{\alpha}
\renewcommand{\b}{\beta}
\renewcommand{\c}{\gamma}
\renewcommand{\d}{\delta}
\newcommand{\e}{\epsilon}

\newcommand{\shalf}{\frac{1}{2}}
\newcommand{\pa}{\partial}

\newcommand{\ra}{\rangle}
\newcommand{\lan}{\langle}
\newcommand{\nn}{\nonumber \\}

\begin{document}

\topmargin 0pt
\oddsidemargin 5mm
\def\bbox{{\,\lower0.9pt\vbox{\hrule \hbox{\vrule height 0.2 cm
\hskip 0.2 cm \vrule height 0.2 cm}\hrule}\,}}
\newcommand{\NP}[1]{Nucl.\ Phys.\ {\bf #1}}
\newcommand{\PL}[1]{Phys.\ Lett.\ {\bf #1}}
\newcommand{\CMP}[1]{Comm.\ Math.\ Phys.\ {\bf #1}}
\newcommand{\PR}[1]{Phys.\ Rev.\ {\bf #1}}
\newcommand{\PRL}[1]{Phys.\ Rev.\ Lett.\ {\bf #1}}
\newcommand{\PTP}[1]{Prog.\ Theor.\ Phys.\ {\bf #1}}
\newcommand{\PTPS}[1]{Prog.\ Theor.\ Phys.\ Suppl.\ {\bf #1}}
\newcommand{\MPL}[1]{Mod.\ Phys.\ Lett.\ {\bf #1}}
\newcommand{\IJMP}[1]{Int.\ Jour.\ Mod.\ Phys.\ {\bf #1}}
\newcommand{\CQG}[1]{Class.\ Quant.\ Grav.\  {\bf #1}}
\def\G{\Gamma}
\def\D{\Delta}
\def\ve{\varepsilon}
\def\z{\zeta}
\def\t{\theta}
\def\vt{\vartheta}
\def\i{\iota}
\def\r{\rho}
\def\vr{\varrho}
\def\k{\kappa}
\def\l{\lambda}
\def\L{\Lambda}
\def\o{\omega}
\def\O{\Omega}
\def\s{\sigma}
\def\vs{\varsigma}
\def\S{\Sigma}
\def\vphi{\varphi}
\def\av#1{\langle#1\rangle}
\def\pa{\partial}
\def\na{\nabla}
\def\hg{\hat g}
\def\un{\underline}
\def\ov{\overline}

\begin{titlepage}
\setcounter{page}{0}

\begin{flushright}
COLO-HEP-357, OU-HET 209 \\
hep-th/9504033 \\
March 1995
\end{flushright}

\vspace{5 mm}
\begin{center}
{\large Thermodynamics of Quantum Fields in  Black Hole Backgrounds}
\vspace{10 mm}

{\large S. P. de Alwis\footnote{e-mail: dealwis@gopika.colorado.edu}
}\\
{\em Department of Physics, Box 390,
University of Colorado, Boulder, CO 80309}\\
\vspace{5 mm}
{\large N. Ohta\footnote{e-mail: ohta@phys.wani.osaka-u.ac.jp}}\\
{\em Department of Physics, Osaka University,
Toyonaka, Osaka 560, Japan}\\

\end{center}
\vspace{10 mm}

\centerline{{\bf{Abstract}}}
We discuss the relation between the micro-canonical and the canonical
ensemble for black holes, and highlight some problems associated with
extreme black holes already at the classical level. Then we discuss
the contribution of quantum fields and demonstrate that the partition
functions for scalar and Dirac (Majorana) fields in static space-time
backgrounds, can be expressed as  functional integrals in the
corresponding optical space, and point out that the difference
between
this and the functional integrals in the original metric is a
Liouville-type action. The optical method gives both the correction
to the black hole entropy and   the bulk contribution to the entropy
due to the radiation, while (if the Liouville term is ignored) the
conical singularity method just gives the divergent contribution to
the black hole entropy. A simple derivation of a general formula for
the free energy in the high-temperature approximation is given and
applied to various cases. We conclude with a discussion of the second
law.

\end{titlepage}
\newpage
\renewcommand{\thefootnote}{\arabic{footnote}}
\setcounter{footnote}{0}

\setcounter{equation}{0}
\sect{Introduction}
\indent
There has been renewed interest in the entropy of the black holes
and its connection with the puzzle of information loss. If a black
hole is formed from a pure quantum state and decays by emitting
Hawking radiation in a thermal state in accordance with Hawking's
``semi-classical" argument~\cite{Haw}, it appears that a pure state
evolves into a mixed state in contradiction to one of the fundamental
laws of quantum mechanics, and quantum mechanical information is
lost. This is quantified by the entanglement entropy of the final
(mixed) state. There is also the related (though not identical)
question of the validity of the second law of
thermodynamics in semi-classical black hole physics.

The two issues mentioned above are actually related to two different
concepts of entropy. What is relevant for the second question is the
{\it thermodynamic or coarse grained entropy}, and applies even for
classical black holes. This is associated with a description of
the system that only specifies its macroscopic characteristics.
It is this entropy that is expected to satisfy the second law of
thermodynamics and is given according to Boltzmann by the
logarithm of the available phase space.\footnote{As usual the
ultraviolet catastrophe has to be regulated by introducing $\hbar$;
i.e. the Boltzmann entropy of a system for which the available phase
space volume is $\G$ has the value $S=\ln{\G\over{h^{N}}}$, N being
half the dimension of the phase space.}
This concept can be extended to quantum mechanics where the
phase space volume (or rather the number of phase space cells) would
have to be replaced by the dimension of the Hilbert space available
to the system (the two being the same in the semi-classical limit).
However it is difficult to understand how the Bekenstein-Hawking
entropy of a black hole~\cite{bek,hawk} which is proportional to
the area of the global horizon $(S_{BH}={A\over 4 G_{N}\hbar}$)
could be interpreted as the Boltzmann entropy since the latter is
expected to be an extensive quantity.\footnote{See ref. \cite{thzu},
however.}

On the other hand, in quantum mechanics there is another concept of
entropy which is applicable when a  system is described
in terms of a set of (commuting) dynamical variables which are not
complete. Typically this arises when the object under study is part
of a larger system (which may be in a pure state) and therefore has
to be described by a density matrix $\r$ obtained by tracing over
all the variables of the rest of the large system.
The microscopic entropy is then defined to be $S=-tr\r\ln\r$. This
entropy arises because of the correlations between the states of
the object and the states of the environment. In quantum
field theory this entropy may be associated with the fields inside a
region of space which are correlated with the fields outside.
The system is then the field configuration in the region in question
and the term {\it geometric entropy} has been used for
its microscopic entropy.
It has been argued \cite{bom,sred} that (at least in flat space) the
geometric entropy is proportional to the area. This is plausible
because of the result that the entropy in the interior of the region
is exactly equal to the entropy of the outside so that each must be
dependent on the common boundary \cite{sred}. Clearly if these
arguments can be extended to black hole space-time, then one could
have a statistical mechanics basis for the Bekenstein-Hawking
entropy at least in the case where the Einstein term is induced by
quantum fluctuations like in string theory. Some steps in this
direction  have recently appeared in the literature \cite{suss,call}.
In particular, a functional integral expression for the geometric
entropy of Rindler space has been given in \cite{call}. On the other
hand, Ref.~\cite{suss} gives two different calculations of the
Rindler entropy.\footnote{In all of these calculations the entropy
per unit area on the Rindler horizon is
a constant in agreement with the expectations for geometric entropy.}
One is a direct thermodynamic (Hamiltonian) calculation of
the thermal entropy of a gas of bosons in Rindler space. The other is
a calculation of the path integral representation of the thermal
partition function as in \cite{call} . The calculations of the
entropy in both these cases agree. However the free energy has
different values. The path integral expression is of course the same
in both \cite{suss} and \cite{call} and both give a vanishing free
energy at the Rindler temperature. On the other hand, the direct
thermodynamic calculation gives a non-zero value. One of the purposes
of this work is to investigate this difference. It will
be found that it arises from the different measures that one has to
use for the thermal partition function and the geometric one.

The relevance of these works to the black hole is that the Rindler
space is the limit of infinite black hole mass in a Schwarzschild
space-time. However it is clearly desirable to study the finite mass
case. In particular one would like to distinguish between the two
types of entropy. We expect that the geometric entropy is
proportional to the area of the horizon and there should be no bulk
term. On the other hand, the thermal entropy should presumably have
a bulk term corresponding to the gas of particles (quantum fields)
outside the black hole as well as an area term, i.e. a correction to
the black hole entropy.  Now if one considers a
series of quasi-static frames corresponding to
black holes of decreasing mass, one should find that the microscopic
entropy decreases to zero. This entropy is divergent because of
correlations between fields which are arbitrarily close to
the horizon on either side of it \cite{thooft}. Thus this procedure
requires one to impose a cut off at the horizon that is kept fixed in
every frame. Alternatively one has to consider a theory with a
natural short distance cutoff (e.g. string theory) as advocated by
Susskind and Uglum \cite{suss}. If the area law  is
a measure of the information hidden behind the horizon,
clearly it must come out in the Hawking radiation as the black
hole decays. Thus it seems very important to establish the area law
for the geometric entropy in the case of a finite mass black hole.
Unfortunately there is no straightforward generalization
of the Rindler space arguments of~\cite{call}.

We consider next the canonical ensemble for quantum fields in a given
static background. It is shown that what one obtains for the
partition function is a Euclidean path integral but the relevant
metric is not the original Rindler or black hole one but is
the so-called optical metric introduced by Gibbons and Perry
\cite{gp}.\footnote{This metric was  used in connection
with this problem many years ago by Dowker and Kennedy~\cite{dk} who
also seem to have been the first to derive the high-temperature
asymptotic expansion given below (3.26). However these authors did
not discuss the divergences in the free energy and entropy arising
from the divergence of the optical volume. The metric has also been
used recently by \cite{barb,emp}. However there was no discussion
of the relation of the optical method to the conical singularity
one in the original metric, in any of these papers.}
The optical metric is conformally related to the original (Rindler
or black hole) metric and our expression differs from those
in the geometric formulation  by an action analogous to the Liouville
action. For instance in  two-dimensional Rindler space, we show that
this partition function is the same as that considered by Callan and
Wilczek~\cite{call}
except for the well-known Liouville action. It is this Liouville term
which gives a non-zero free energy to Rindler space (at the Rindler
temperature $T=1/2\pi$). In the four-dimensional case (even for
Rindler space) the partition function cannot be calculated exactly.
However in any dimension and for any static backgrounds, we can
obtain a formula for the free energy (and
hence the entropy) in the high-temperature approximation.

Unlike what one expects for the microscopic entropy, for our
thermodynamic entropy one obtains also a bulk contribution
representing the free energy of a gas of particles as well as terms
which depend on the mass of the black hole and which would go to
zero as  the black hole decays. These terms enable us to discuss the
operation of the second law.
It should be stressed again that this applies to the thermodynamic
entropy and not to the entanglement (geometric) entropy.

In four-dimensional black hole backgrounds, we find that in addition
to the linear divergence in the entropy (and free energy), there is
also a logarithmic divergence. Thus renormalization of $G_N$ alone
is not sufficient to make the entropy finite. One also needs to
introduce a bare $R^2$ term to the original action to deal with
this.\footnote{This has already been pointed out in \cite{solod}.}
We also investigate the entropy of quantum fields around
Reissner-Nordstr\"om and dilaton black holes. In the former case,
we find that there is a cubic order divergence as the extreme limit
is approached. This is present in addition to the linear (quadratic
in terms of the proper cutoff) and logarithmic divergences of the
Schwarzschild case which can be incorporated into the renormalization
of Newton's constant \cite{suss} and coefficients of higher
derivative terms in the effective action. We believe that this
signifies a break down of the thermal ensemble for these extreme
holes. For the dilaton black hole, on the other hand, we find that
the free energy has a linear as well as a logarithmic divergence.
The former is zero (as is the ``classical" entropy), but the latter
is non-zero in the extreme limit. A short account of these
calculations was presented by us in \cite{do}. The calculations for
the dilaton and Reissner-Nordstr\"om black holes have also been done
by a different method in~\cite{gm}, with  results in agreement
with ours.

\sect{The ``Classical" Entropy of Black Holes}
\indent
We begin with the micro-canonical formulation of (quantum)
statistical mechanics.\footnote{The micro-canonical ensemble for
black holes is discussed in some detail in \cite{by}.} Introducing
the Hamiltonian operator $H$ of the system and
its eigenstates $|E,r\ra$ ($H|E,r\ra=E|E,r\ra$) where $r$ labels the
eigenvalues of all other commuting observables which serve to
characterize the quantum state. The degeneracy of the energy
eigenstates then gives us a definition of the entropy. Writing
$N(E) =tr\d (E-H)$, we define the entropy as the logarithm of this
degeneracy,
\EQ
\label{entropy}
S=\ln N(E).
\EN
The (inverse) temperature is then defined by
\EQ
\label{temperature}
T^{-1} ={\pa S\over \pa E}.
\EN

In taking this formalism over to discuss the thermodynamics of
space-time itself, the first problem one faces is the definition of
energy. As is well known, one may define an energy when the
space-time
is asymptotically flat. Thus we define the quantum mechanics of
metrics and other fields by (say) a functional integral formulation
in which the fields die away asymptotically (in the space variables).

In order to calculate this entropy in quantum field theory, one needs
to introduce the Laplace transform of the degeneracy $N(E)$, i.e. the
so-called partition function:
\EQ
\label{partfn}
Z(\b)=\int dE e^{-\b E}N(E) =\int dE e^{-\b E} tr \d (E-H)=tr e^{-\b
H}.
\EN
It should be stressed that here $\b$ is just the Laplace transform
variable and we do not necessarily have to give it a physical
interpretation as an inverse temperature. The latter will be the case
in a rather different context where we have the canonical
ensemble with the system being in equilibrium with a heat bath.

By standard arguments, we obtain
\EQ
Z(\b )=\int_{\b} [dg][d\phi ]e^{-I_g^{\b}-I_m^{\b}},
\EN
where
\EQ
I_g^{\b} = -{1\over 16\pi}\int_{\cal M_{\b}}R - {1\over
8\pi}\int_{\pa \cal M} K,
\EN
and $I_m^\b$ are the Euclidean gravitational and matter actions over
a space-time manifold $\cal  M_{\b}$ which has toroidal topology in
the (Euclidean) time direction with period $\b$. The subscript on the
functional integral is an instruction to impose these boundary
conditions on the fields. In the next section, it will be shown that
the measure in the functional integration is not
the naive measure but this fact is irrelevant for our ``classical"
considerations in this section.

In order to evaluate this in the saddle point approximation, one
needs the classical action. This has the value~\cite{hawk,teit}
(assuming that the dominant contribution to the functional integral
coming from the matter action is the zero field one)
\EQ
\label{classaction}
I=\b M-{A\over 4}=\b M-4\pi M^2.
\EN
The first term is simply the value of the classical Hamiltonian
multiplied by the Euclidean time interval $\b$,
and the second term is quarter the area of the horizon for black hole
solutions, which we have set equal to its value for the Schwarzschild
solution in the second equality. $M$ of course is an integration
constant that is equal to the value of the boundary Hamiltonian i.e.
the mass of the space-time. Now usually in the literature this value
is substituted into the functional integral, thus approximating it by
its classical value of the integrand at some fixed $M$. However it is
more appropriate to integrate over $M$ since one has an integral over
all metrics. Indeed that is precisely what one should expect from
a partition function i.e. a function of a variable $\b$ that is
conjugate to the energy! Thus our approximate evaluation of the
partition function gives, (restricting ourselves to Schwarzschild
spaces)
\EQ
\label{appart}
Z\simeq\int dM e^{-\b M+4\pi M^2}.
\EN

This integral is of course divergent but the important point is
that this divergence has a physical interpretation. Comparing with
(\ref{partfn}), we see that $N(E)=e^{A\over 4}=e^{4\pi E^2}$ so that
from (\ref{entropy}) and (\ref{temperature})
we get the Bekenstein-Hawking entropy and inverse Hawking temperature
\EQ
S={A\over 4}=4\pi E^2, ~~T_H^{-1} \equiv \b_H=8\pi E.
\EN
The free energy of the space may then be defined in the usual manner
and we get $F=E-T_H S={E\over 2} $.

Similarly for Reissner-Nordstr\"om black holes, $S={A\over 4}=\pi
r_+^2$ where $r_+$ is the radius of the outer horizon
($r_{\pm}=E{\pm}\sqrt{E^2-Q^2}$) and $Q$ is the charge that is
kept fixed as an external macroscopic parameter,
like the energy $E$ of the black hole. From (\ref{temperature}), the
Hawking temperature is then $T ={(r_+-r_-)\over 4\pi r_+^2}$.
In the extreme case $E\to Q$ this tends to  zero and the entropy
$\to \pi E^2$. On the other hand, one may discuss the
thermodynamics of the extreme holes without reference to the
non-extreme case. In this case we have for the action $I =\b M$
(see third paper of \cite{hawk} and \cite{teit}) and the partition
function is
\EQ
\label{ext}
Z=\int e^{-\b M}dM=\b^{-1}.
\EN
Comparing this with (\ref{partfn}), we find $N(E)=1$ i.e. $S=0$.
Hence $T^{-1}={\pa S\over\pa E}=0$. In other words, the Hawking
temperature is infinite! On the other hand, purely geometric
arguments seem to indicate that the Hawking temperature is arbitrary
(see third paper of~\cite{hawk} and \cite{teit}). Basically the
reason is that the topology for the extreme hole is completely
different from that for the non-extreme case and it has no conical
singularity at any temperature. These ambiguities suggest that
the statistical mechanics of these objects is not well-defined.
Further evidence of this will be given in sect.~4 when we
compute the quantum corrections to these classical values.

To summarize then, in this micro-canonical calculation the partition
function need not be given physical significance. It is merely a
calculational device enabling us to use the path integral formulation
of quantum field theory. The corresponding $\b$ is just the Laplace
transform variable and need not be interpreted as the inverse
temperature of the system. The entropy and physical inverse
temperature are calculated from (\ref{entropy},\ref{temperature}),
and yield the well-known Bekenstein-Hawking and Hawking results,
respectively.

The canonical emsemble, on the other hand, is not well-defined for
black holes since the partition function is divergent. Note that
unlike the partition function for strings which diverges only above
the Hagedorn temperature, in our case it is divergent at all
``temperatures" since the degeneracy grows as $e^{E^2}$ rather than
$e^{E}$. Nevertheless one may make some formal arguments. If one
makes the saddle point approximation in (\ref{appart}), we get
$ Z= e^{-{\b^2\over 16\pi}}$. This gives an average value for the
black hole mass $\lan M \ra ={\b\over 8\pi}$. This is the same
relation as in the micro-canonical ensemble but now with the average
mass of the thermal ensemble replacing the actual mass. Thus we have
agreement between the micro-canonical and the canonical ensembles
for the non-extreme Reissner-Nordstr\"om black hole, but for the
extreme case (where $Z$ is actually well-defined) we have from
(\ref{ext}), $\lan M \ra=-{\pa\ln Z\over\pa\b}=\b^{-1}$. Similarly
the entropy evaluated using the canonical ensemble formula gives
$S=-(\b{\pa\over\pa\b}-1)\ln\b=-1+\ln\b$. These relations are not in
agreement either with the results of the micro-canonical calculation
which gave an infinite temperature and zero entropy nor with
the geometric calculation which gave an arbitrary temperature and
zero entropy. There seems to be further evidence that the
thermodynamics of these extreme objects (if it is at all meaningful)
is somewhat strange. Later we will suggest a reason for this peculiar
thermodynamics (see discussion after equation (\ref{extvol})).

\sect{Free Energy of Scalar Fields}
\indent
We wish to derive a functional integral expression for the thermal
ensemble for a free scalar field in a static background with
metric,\footnote{We are ignoring graviton fluctuations. These are
technically more complicated but we do not
expect them to change the qualitative physics.}
\EQ
ds^2= g_{00}dt^2 + h_{ij}dx^idx^j.
\label{metric}
\EN
We write $g=\det g_{\mu\nu}=g_{00}h,~h=\det h_{ij}$ where $\mu ,\nu
=0,...,D-1;~i,j=1,...D-1$. The action is
\bea
S &=&-{1\over 2}\int
d^Dx\sqrt{-g}g^{\mu\nu}\pa_{\mu}\phi\pa_{\nu}\phi
\nn
&=&\int dt\int d^{D-1}x\sqrt h ~[{1\over
2\sqrt{-g_{00}}}\dot\phi^2-{\sqrt{-g_{00}}\over
2}h^{ij}\pa_i\phi\pa_j\phi ].
\label{action}
\ena

The canonical momentum is $\pi={\dot\phi\over \sqrt{-g_{00}}}$ and
the Hamiltonian is
\EQ
H=\int d^{D-1}x{\cal H}=\int d^{D-1}x\sqrt h \sqrt{-g_{00}}
[{1\over 2}\pi^2+{1\over 2} h^{ij}\pa_i\phi\pa_j\phi ].
\label{ham}
\EN
The equal-time canonical commutation relations are
\EQ
[\hat\phi (\vec{x}),\hat\pi (\vec{y})]={i\over\sqrt h}\d
(\vec{x}-\vec{y}).
\EN
We introduce a basis of eigenstates of the field operator $\hat\phi$
which are (delta function) orthonormal and complete on the space of
fields with metric
$||\d\phi ||^2=\int d^{D-1}x {\sqrt h}(\d\phi )^2$. Similarly we may
introduce a basis of eigenstates of the canonical momentum operator
$\pi$. Using the transformation matrix $\langle\phi |\pi
\rangle={1\over\sqrt{2\pi}}e^{i\int d^{D-1}\sqrt{h}\phi\pi}$ we may
then obtain from the usual time-slicing procedure the following
expression for the partition function:
\EQ
Tr [e^{-\b H}] = \int [d\pi]\int_{\phi (0,\vec{x})=\phi (\b,\vec{x})}
[d\phi ]e^{-\int_0^{\b}dt \int d^{D-1}x\sqrt{h}[-i\pi\dot\phi
+{\cal H}]},
\EN
where ${\cal H}=\sqrt{-g_{00}}[{\pi^2\over 2}+{1\over
2}h^{ij}\pa_i\phi\pa_j\phi ]$.

The Gaussian integral over $\pi$ gives the factor
$\Pi_{t,\vec{x}}{1\over (-g_{00})^{1\over 4}} e^{-\int_0^{\b}dt
\int d^{D-1}\sqrt{h}{\dot\phi^2\over 2\sqrt{-g_{00}}}}$
and we finally have the expression
\EQ
\int_{\phi (0,\vec{x})=\phi (\b,\vec{x})}
\prod_{t,\vec{x}}^{}{d\phi\left({h\over g_{00}^E}(t,\vec{x})
\right)^{1\over 4}} e^{-\int_0^{\b}dt \int d^{D-1}x\sqrt{g^E}{1\over
2}g^{E,\mu\nu}\pa_{\mu}\phi\pa_{\nu}\phi},
\label{part}
\EN
where $g^E_{\mu\nu}=(-g_{00},h_{ij})$ is the Euclideanized metric.
Note in the above that although the action in the exponent of the
functional integral is covariant, the measure is not. This is a
reflection of the fact that a particular $3+1$ split is used in
defining the partition function.
Henceforth we will work with the Euclidean metric and the
superscript $E$ will be omitted.

It is convenient to  discuss conformally coupled scalars and to
introduce a mass term, so we will change the matter action
(after partial integration) to
\EQ
S_{\phi}=\int _0^{\b}dt d^{D-1}x\sqrt{g}\phi (K+m^2)\phi ,
\EN
where
\EQ
K\equiv -\bbox + {1\over 4}{D-2\over D-1}R,
{}~\bbox \equiv
{1\over\sqrt{g}}\pa_{\mu}(\sqrt{g}g^{\mu\nu}\pa_{\nu}).
\EN
Thus we may write
\EQ
Tr [e^{-\b H}] = \int_{\phi (0,\vec{x})=\phi (\b ,\vec{x})}
\prod_{t,\vec{x}}^{}{d\phi \O g^{1\over 4} (t,\vec{x})}
e^{-\int_0^{\b}dt d^{D-1}x\sqrt{g}\phi (K+m^2)\phi}.
\label{partition}
\EN

In the above, $\O ={1\over\sqrt {g_{00}}}$ is a conformal factor
which causes a mismatch between the metric background of the action
and that defining the functional integral.  In order to take into
account this mismatch, we use
the optical metric (introduced in \cite{gp} for somewhat different
reasons) and perform a change of field variable. Thus writing
\EQ
\bar g_{\mu\nu}=\O^2 g_{\mu\nu},~~\bar\phi = \O^{2-D\over 2}\phi,
\label{optical}
\EN
we have for the measure
\EQ
\prod_{t,\vec{x}}^{}{d\phi \O g^{1\over 4}
(t,\vec{x})}=\prod_{t,\vec{x}}^{}{d\bar\phi  \bar g^{1\over 4}
(t,\vec{x})}.
\EN
Using also the properties of the Laplacian and the curvature
under a conformal transformation (see for example \cite{bd}), we
finally obtain
\EQ
Tr [e^{-\b H}]=\int_{\bar\phi (0,\vec{x})=\bar\phi (\b ,\vec{x})}
\prod_{t,\vec{x}}^{}{d\bar\phi  \bar g^{1\over 4}
(t,\vec{x})}e^{-\int_0^{\b}dt d^{D-1}x\sqrt{\bar g}\bar\phi (\bar
K+m^2\O^{-2})\bar\phi}.
\label{partopt}
\EN
Note that the optical metric is of the form
\EQ
\bar ds^2=dt^2+{h_{ij}\over g_{00}}dx^i dx^j,
\label{opt}
\EN
and (since the original metric is static) the topology of optical
space is $S^1\times {\cal M}^{D-1}$.

Now much of the recent discussion of black hole entropy has been
carried out using the path integral in the original metric.
So it is important to understand the relation of that to
(\ref{partopt}). Let us take the massless case. Then we have the
well-known result,
\EQ
\int \prod_{t,\vec{x}}^{}{d\bar\phi  \bar g^{1\over 4}
(t,\vec{x})}e^{-\int_0^{\b}dt d^{D-1}x\sqrt{\bar g}\bar\phi \bar
K\bar\phi}= \int \prod_{t,\vec{x}}^{}{d\phi g^{1\over 4}
(t,\vec{x})}e^{-\int d^{D}x\sqrt{ g}\phi K\phi + \G [\O,g]},
\EN
where $\G$ is the Liouville-type action. In two dimensions this
has the form~\cite{poly}
\EQ
\label{2dliouville}
\G [\O ,g]={1 \over 24\pi}\int{\sqrt g} g^{\mu\nu}
 \pa_\mu \s \pa_\nu \s,
\EN
while in four dimensions it is~\cite{rieg}
\EQ
\label{4dliouville}
\G =\int d^4x\sqrt{g}[2 b' \s\D_4\s +\{bF +b'(G-{2\over 3}
\bbox R)\}\s],
\EN
where $\s =\ln\O$, $F$ is the square of the Weyl tensor, $G$ is
the Euler density, $\D_4 $ is a fourth order differential operator
which also involves the Ricci tensor and scalar curvature of $g$.
In eq.~(\ref{4dliouville}), $b,b'$ are numerical coefficients whose
values we do not need for the moment. Thus we have two different
expressions for the free energy at an inverse temperature $\b$:
\bea
-\b F&=&-{1\over 2}\ln\det [K_{\b}+m^2]+\G [\O,g]\nn
&=&-{1\over 2}\ln\det [\bar K_{\b}+m^2\O^{-2}] .
\label{determinant}
\ena

The first line is the calculation in the original metric while the
second line gives the optical metric version. The difference between
the two determinants is the Liouville-type action (upto terms which
vanish as
$m\rightarrow 0$).
The determinant on the first line has the heat kernel representation
\EQ
\ln\det [K_{\b} +m^2]= -\int_{\e}^{\infty} {ds\over s}\int\sqrt g
d^Dx H(s|x,x),
\EN
where the heat kernel is defined by $H(s|x,x') =e^{-s (K+m^2)}{1\over
\sqrt{g}}\d^{D}(x-x')$ and $\e$ is an ultraviolet cutoff.
The trace of the heat kernel has the asymptotic expansion (see for
example~\cite{bd}) \footnote{When the metric has conical
singularities the coefficients $B_i (i\ge 1)$ will acquire additional
surface terms at the horizon which vanish as $T\to T_H$~\cite{furs}.
Since we are not going to work with this formulation we will
omit writing these out.}
\EQ
\int\sqrt g d^Dx H(s|x,x) = (4\pi s)^{-{D\over 2}}e^{-sm^2}
\sum_{k=1}^{\infty}{(-s)^k\over k!}B_k,
\label{heatkerntrace}
\EN
where
\EQ
B_0=\int_{\cal M}\sqrt g, ~~B_1=(\xi -{1\over 6})\int_{\cal M}\sqrt
gR,
\label{bzero}
\EN
(with $\xi ={1\over 4}{D-2\over D-1}$)
and the general term has the structure
\bea
B_k =\int_{\cal M}\sqrt g [R\bbox^{k-2}R&+&\sum_{0\le i\le
2k-6}R\nabla^iR\nabla^{2k-6-i} R +...\nn &+&\sum_{0\le i\le
k-3}R^i(\nabla R)R^{k-i-3}\nabla R+R^k].
\label{bkay}
\ena
On dimensional grounds, it is clear that only the $B_0$ term will
give a bulk contribution which however diverges like $\e^{-D}$.
This term is independent of the temperature and does not contribute
to the free energy nor to the entropy.\footnote{This observation has
been made earlier by B. Allen~\cite{allen} who considered the
difference between the partition function for the normal ordered
hamiltonian and the optical space~\cite{gp} functional integral.
However the relation between the partition function and optical
method on the one hand, and the functional integral in the original
metric on the other, was not discussed there.} It should be canceled
against the bare cosmological constant in order that the original
(e.g. Schwarzschild instanton) solution may be a valid saddle point.
In the corresponding supersymmetric case this contribution would be
zero. The entropy comes solely from the curvature terms in the
determinant since a nontrivial $\b$-dependence can arise only from
the conical singularity $\int R\sim 2\pi -\b$ that is introduced when
one moves away from the Hawking temperature in order to compute the
entropy~\cite{teit,suss,call}. Since in Rindler space the curvature
at the Hawking temperature is exactly zero, these $\b$-dependent
terms vanish at the Hawking temperature. Consequently the entire
free energy of the gas of particles at the Hawking temperature must
come from the  Liouville-type action (or from non-local terms which
are not modeled by the asymptotic expansion). However except in two
dimensions the Liouville-type term does not appear to be infrared
divergent. We must thus conclude that there is no bulk free energy
in Rindler space or that it must arise from terms which do not occur
in the asymptotic expansion. Furthermore
in this method the origin of the bulk  entropy of the  gas of bosons
(in any static background) is also obscure. Again the local terms
occurring in the asymptotic expansion of the heat kernel cannot
contribute to this bulk entropy since none of them (except the
cosmological constant term which cannot contribute to the entropy)
can have a cubic infrared divergence. On the other hand this
contribution does not seem to  come from the Liouville-type
action term in the first line of (\ref{determinant}) (see for
instance the expression (\ref{4dliouville}) for four dimensions).
It must thus come from non-local terms in the determinant. In the
case of Rindler space we will confirm (in sect. 5) from the optical
method that both the free energy and entropy are proportional to
the area, i.e. there is no bulk term.

The presence of the Liouville-type term makes calculation of entropy
using the original metric somewhat awkward. Since the Liouville-type
action involves curvature terms (see (\ref{4dliouville})), there
will be contributions from it to the entropy and a nontrivial
$\b$-dependence will be introduced through the conical singularities.
However these terms will be finite so one can safely calculate all
the {\it divergent} contributions to the entropy by ignoring the
Liouville-type action.

Let us now turn to the optical space method. In the second line of
(\ref{determinant}), the
relevant manifold has the topology of $S^1\times {\cal M_{\rm D-1}}$
and the heat kernel factorizes since
$-\bar K_\b=\pa_{\o}^2+\bar\bbox_{D-1} -\xi R$ and $\bar \bbox_{D-1}$
is independent of $\o$. Thus we have
\EQ
-\b F ={1\over 2}\int {ds\over s}\int_0^{\b}d\o H_{S^1}
\int\sqrt{\bar g} d^{D-1}x \bar H_{D-1}(s|x,x),
\label{freeen}
\EN
where
\EQ
H_{S^1}=\b^{-1}\sum_{n=-\infty}^{\infty}e^{-s({2n\pi\over
\b})^2}={1\over (4\pi s)^{1\over 2}}\sum_{n=-\infty}^{\infty}
e^{-{\b^2n^2\over 4s}},
\label{hbeta}
\EN
after using the Poisson resummation formula. In the second integral
of (\ref{freeen}), we may use the asymptotic expansion of the heat
kernel given by (\ref{heatkerntrace})-(\ref{bkay}) except that
$\sqrt g$ is replaced by $\sqrt {\bar g }$, $e^{-sm^2}$ by
$e^{-s\O^{-2}m^2}$ (which goes inside the integrals defining the
$B$'s), $R $ by $\bar R$ and $\cal M_D$ by $\cal M_{\rm
D-1}$.\footnote{There are also $m$-dependent non-covariant terms
coming from the non-commutativity of $\O$ and the Klein-Gordon
operator. However we ignore them since we are interested only in
the high-temperature limit where mass terms are irrelevant. We keep
the mass-dependent exponential factor merely as an infrared
regulator until we finally take the high-temperature limit.}

Thus one gets from the optical
point of view, the following expression for the
free energy after subtracting the zero temperature cosmological
constant term (i.e. the $n=0$ term in the thermal sum):
\EQ
F(\b )=-{1\over 2}\int_0^{\infty}{ds\over s} {1\over (4\pi s)^{D\over
2}}\sum_{n\ne 0}
e^{-{\b^2n^2\over 4s}}\sum_{k=0}^{\infty}{(-s)^k\over k!}\bar B_k.
\label{freeenergy}
\EN

It is important to note that in this calculation the free energy
indeed has as expected an ultraviolet divergence, but it comes
from the divergence of the optical metric and not from the $s=0$
end of the proper time integral. In fact the only divergence
comes from the divergence of the density $\sqrt{\bar g}$ at the
horizon. To clarify this and several other issues we will look at
the examples of Rindler and black hole spaces in the following
sections. Before we do that, let us close this section
by writing down the high-temperature limit of the above formula for
the free energy. This is easily evaluated by first changing the
variable of the proper time integral from $s$ to $u=\b^{-2}s$ and
then neglecting the higher powers of $\b^2$ coming from the expansion
in (\ref{freeenergy}). Thus we get
\EQ
F(\b )=-T^D\int{du\over u}{1\over (4\pi u)^{D\over 2}}
\sum_{n=1}^{\infty}e^{-{n^2\over 4u}}\int_{ { \cal M_{\rm D-1}}}
\sqrt{\bar g}e^{-{\O^{-2}m^2\over T^2}}.
\label{hightemp}
\EN
Or if our system consists of only particles with masses that are
small compared to the temperature (the above form is suitable for a
generalization to string theory where this is certainly not the
case), we get
\bea
F&=&-T^DV_{D-1}\int_0^{\infty}{du\over u}{1\over (4\pi
u)^{D\over2}}\sum_{n= 1}^{\infty}e^{-{n^2\over 4u}}\nn
&=& -{T^D V_{D-1}\over {\pi^{D\over 2}}} \Gamma\left({D\over
2}\right)
\zeta(D),
\label{highfree}
\ena
where
$V_{D-1}=\int_{ { \cal M_{\rm D-1}}}\sqrt{\bar g}$ is the
volume of optical space. This is just the free energy of a gas of
(massless) particles in a box whose volume is given by the optical
measure. Thus in four dimensions we have
\EQ
F=-{\pi^2\over 90}V_3 T^4,~~S={2\pi^2\over 45}V_3T^3,
\label{exam}
\EN
and in two dimensions
\EQ
F=-{\pi\over 6}V_1 T^2,~~S={\pi\over 3}V_1T.
\label{exam2}
\EN
Calculating the free energy and hence the entropy of the scalar
fields in different backgrounds is now just a triviality. One just
plugs in the volume of optical space corresponding to each metric.

\sect{Free Energy of Fermions}
\indent
In this section, let us briefly show how our above results are
modified for Dirac (Majorana) fermions. This will be important
especially when we consider supersymmetric case. The contribution
of fermions was also discussed in Ref.~\cite{wil}, but our unified
treatment is much simpler.

We introduce the vierbein fields of the form
\EQ
v_\mu^\a = \left[
\begin{array}{cc}
v_0^0 & 0 \\
0 & w_i^a
\end{array}
\right], \qquad
v_\a^\mu = \left[
\begin{array}{cc}
(v_0^0)^{-1} & 0 \\
0 & w_a^i
\end{array}
\right],
\EN
where $\mu=0,1,\cdots,D-1;\; i=1,2,\cdots ,D-1$ are curved indices
and $\a=0,1,\cdots,D-1;\; a=1,2,\cdots ,D-1$ are tangent space ones.
The action is
\bea
S&=& \int d^D x v ( i{\bar \psi}\c^\a v_\a^\mu \nabla_\mu \psi
 - m {\bar\psi}\psi ) \nn
&=& \int dt\int d^{D-1} x w ( i\psi^\dagger \nabla_0 \psi
 +i{\bar \psi}\c^a v_0^0 w_a^i \nabla_i \psi-m v_0^0{\bar\psi}\psi),
\ena
where $v=\det v^\a_\mu=v_0^0 w,\ w=\det w^a_i, \nabla_\mu=\pa_\mu
+\G_\mu$ with the connection $\G_\mu=\shalf \S^{\a\b}
v_\a^\nu \frac{\pa}{\pa x^\mu}v_{\b\nu}$, $\S^{\a\b}$ being the
generator of the Lorentz group for a Dirac field.

The canonical momenta are $\pi_\psi=i \psi^*$ and
the Hamiltonian is
\EQ
H=\int d^{D-1}x{\cal H}=
\int d^{D-1}x v (-i\psi^\dagger(v_0^0)^{-1} \G_0 \psi
-i {\bar\psi}\c^a w_a^i\nabla_i\psi +m{\bar\psi}\psi).
\EN

Introducing the time-slicing procedure to define path integral, as
in the scalar fields, we find the partition function is given by
\EQ
Tr[e^{-\b H}] = \int_{\psi (0,\vec{x})=-\psi (\b,\vec{x})}
\prod_{t,\vec{x}}^{} \left(d\psi d\psi^* w \right)
e^{-\int_0^{\b}dt \int d^{D-1}x v {\cal L}},
\EN
where
\EQ
{\cal L}= (i{\bar \psi}\c^\a v_\a^\mu \nabla_\mu \psi
 -m {\bar\psi}\psi).
\EN

Introducing the optical vierbein
\EQ
{\tilde v}_\mu^\a=\frac{v_\mu^\a}{v_0^0}\equiv \O v_\mu^\a ,
\EN
and ${\tilde\psi}=\Omega^{(1-D)/2}\psi \;({\tilde \psi^*}
=\Omega^{(1-D)/2}\psi^*)$, we find\footnote{Here we ignore the mass
terms for simplicity. It is easy to recover them in the following
formulae and in any case they drop out in our high-temperature
approximation.}
\bea
Tr[e^{-\b H}] &=&
\int \prod_{t,\vec{x}}^{}\left({\tilde v}d{\tilde\psi}
d{\tilde{\psi^*}}
\right) e^{-\int_0^{\b}dt \int d^{D-1}x {\tilde v} {\cal L}},\nn
&=& \int \prod_{t,\vec{x}}^{} \left( v d\psi d{\psi^*}\right)
e^{-\int_0^{\b}dt \int d^{D-1}x v {\cal L} + \G_F},
\label{pat}
\ena
where again $\G_F$ is a Liouville-type action for fermions.

We calculate the functional determinant resulting from the first
line of (\ref{pat}) using the heat kernel. The result can be
expressed as
\EQ
\b F= \shalf \zeta'_F(0),
\EN
where
\EQ
\zeta_F(p)=\sum_n \frac{1}{\Gamma(p)}\int_0^\infty ds s^{p-1}
\int d^{D-1}x \sqrt{\bar g}\lan x|e^{-s {\bar K}_F} |x\ra,
\EN
with ${\bar K}_F$ is the Feynman propagator in the optical vierbein.
Noting the antiperiodic conditions for fermions and using the
Poisson resummation formula, we get
\EQ
\zeta_F(p) = \frac{\b}{\Gamma(p)}\int_0^\infty ds s^{p-1}
\frac{1}{(4\pi s)^\frac{D}{2}} \sum_{n=-\infty}^{\infty}(-1)^n
e^{\frac{\b^2n^2}{4s}} \sum_{k=0}^\infty \frac{(-t)^k}{k!}\bar
B_k^{(F)},
\EN
where use has also been made of the asymptotic expansion of the heat
kernel for fermions. The coefficients $\bar B_k^{(F)}$ are given by
(see for example~\cite{bd})
\EQ
\bar B_0^{(F)}= f \int_{\cal M}\sqrt{\bar g},
{}~\bar B_1^{(F)}= -{ f\over 12} \int_{\cal M}\sqrt{\bar g}\bar R,
\label{bone}
\EN
etc. with $f=2^{[D/2]}$ being the number of spinor components
in $D$ dimensions. Thus after dropping the $n=0$ term which
can be absorbed into the zero-temperature cosmological constant (in
a supersymmetric case it will cancel the bosonic contribution),
we finally obtain the free energy
\EQ
F(\b )={1\over 2}\int_0^{\infty}{ds\over s} {1\over (4\pi s)^{D\over
2}}\sum_{n\ne 0} (-1)^n e^{-{\b^2n^2\over 4s}}
\sum_{k=0}^{\infty}{(-s)^k\over k!}\bar B_k^{(F)},
\label{fermienergy}
\EN
Note the close similarity to eq.~(\ref{freeenergy})
for scalar field.

In the high-temperature limit, we can use a similar procedure for
scalar fields to obtain
\bea
F&=&T^DV_{D-1}\int_0^{\infty}{du\over u}{f\over (4\pi
u)^{D\over2}}\sum_{n= 1}^{\infty}(-1)^n e^{-{n^2\over 4u}}\nn
&=& -2^{[D/2]}(1-2^{1-D}) {T^D V_{D-1}\over {\pi^{D\over 2}}}
 \Gamma\left({D\over 2}\right) \zeta(D).
\label{highfermi}
\ena
where $V_{D-1}$ is again the volume of optical space.

In four dimensions we have
\EQ
F=-\frac{7}{2} {\pi^2\over 90} V_3 T^4,~~
S=\frac{7\pi^2}{45} V_3 T^3,
\label{exam3}
\EN
and in two dimensions
\EQ
F=- {\pi\over 6} V_1 T^2,~~
S=\frac{\pi}{3} V_1 T.
\label{exam4}
\EN
If the fermions are Majorana, these get an additional factor 1/2.

In the following sections we will apply these formulae to the
examples of Rindler and black hole spaces.

\sect{Rindler Space}
\indent
Euclidean Rindler space has the metric
$ds^2=R^2d\o^2+dR^2+dx_{\bot}^2$. To avoid a conical singularity at
the origin $\o$ must be identified with period
$\b=2\pi$ corresponding to a temperature $T={1\over 2\pi}$.
The logarithm of the partition function should then give us the
free energy of a gas of bosons at this temperature (multiplied  by
$-2\pi$).
If one used the covariant functional integral to evaluate the
partition function, one can just transform from polar coordinates
to cartesian coordinates to demonstrate that the free energy is zero.
Of course the functional integral will have the usual divergence
associated with the zero temperature cosmological constant which must
be canceled against the bare constant. Moreover in
the supersymmetric case,  this will be strictly zero though at
any finite temperature one must get a non-zero free energy due to
the different boundary conditions on bosons and fermions. Thus
the path integral with the original metric does not give the right
free energy, and the reason is that the correct path integral
for the evaluation of the thermal ensemble is (\ref{partopt}) and not
the one with the original metric. On the other hand, in the
evaluation of the entropy one needs to go away from the Rindler
temperature, thus introducing a conical singularity at the origin
\cite{teit,suss,call} when working with the original metric. The
value of the entropy agrees with that computed in the thermal
ensemble. As we will see (at least in two dimensions),
this curiosity can be explained precisely in terms of the well-known
Polyakov term.

Rindler space (with $\b = 2\pi$) is flat but
the corresponding optical space which has metric
$\bar{ds^2}=d\o^2+{1\over R^2}(dR^2+dx_{\bot}^2)$ has catuuuure
$\bar {\cal R}=-(D-1)(D-2)$ which however vanishes for $D=2$.
Thus in this case we have for the thermal partition function
the expression (we will set $m=0$ for simplicity. It is irrelevant
anyway in the high-temperature limit):
\bea
Tr [e^{-\b H}]
&=&\int_{\bar\phi (0,\vec{x})=\bar\phi (\b ,\vec{x})}
\prod_{\o ,\vec{x}}^{}{d\bar\phi  \bar g^{1\over 4}
(\o ,\vec{x})}e^{-\int_0^{\b}d\o d^{D-1}x\sqrt{\bar g}
{\bar g}^{\mu\nu}\pa_{\mu}\bar\phi\pa{\nu}\bar\phi} \nn
&=& \int_{\phi (0,\vec{x})=\phi (\b ,\vec{x})}
\prod_{\o ,\vec{x}}^{}{d\phi
g^{1\over 4} (\o ,\vec{x})}e^{-\int _0^{\b}d\o dR\sqrt{ g}
{g^{\mu\nu}}\pa_{\mu}\phi\pa{\nu}\phi +\G } .
\label{twodrindler}
\ena
In the above $\G ={1\over 48\pi}\int d^2x\sqrt{g}{1\over 2}g^{\mu\nu}
\pa_{\mu}\ln\O^2 \pa_{\nu}\ln\O^2$ is the Liouville action coming
from the two-dimensional conformal anomaly. Let us first use the
second line in the above equation to evaluate this. For $2D$ Rindler
space $\O ={1\over R}$. Hence since the scalar field action gives no
contribution to the free energy at $\b =2\pi$, we have from
the Liouville action\footnote{For a related calculation see
\cite{gid}.}
\EQ
-2\pi F={1\over 48\pi}\int_0^{2\pi}d\o \int_{\e}^{L} RdR {1\over
R^2},
\EN
where we have regulated the integral by introducing cutoffs at both
short and long distances. Thus we get a free energy
\EQ
F=-{1\over 24\pi}\ln{L\over\e },
\label{freeboson}
\EN
for a single scalar field, whereas for a Dirac spinor similar
procedure yields the same result. (If the fermion is Majorana,
the result is half this value.)
This is in agreement with the direct Hamiltonian calculation of the
free energy of a 1-dimensional gas of free bosons at a temperature
$2\pi$. To obtain the entropy one has to differentiate with respect
to $\b$ and going away from the Rindler value $\b =2\pi $ introduces
a conical singularity which gives (from the curvature terms in
the evaluation of the determinant) a $\b$-dependence.
There is no contribution to the entropy from the Liouville term
since it gives only a $\b$-independent contribution to $ F$.

It is instructive to look at the calculation also from the optical
space path integral (first line of (\ref{twodrindler})). In this case
after changing the spatial variable from $R$ to $x=\ln R$, we have
\EQ
e^{-\b F}=  \int_{\phi (0,\vec{x})=\phi (\b ,\vec{x})}
\prod_{\o ,\vec{x}}^{}{d\phi  (\o ,\vec{x})}e^{-\int _0^{\b}d\o
\int_{-\infty}^{\infty} dx\bar\phi (\pa_{\o}^2+\pa_x^2)\bar\phi}.
\EN
This is the usual path integral form for the partition function
for a gas of free bosons at an inverse temperature $\b$.
Indeed, from eq.~(\ref{exam2}) and eq.~(\ref{exam4}), we get
(\ref{freeboson}) for $T=1/2\pi$.
Thus the optical metric calculation directly gives us the correct
expression for the free energy as well as the entropy of Rindler
space.

Rindler space is more complicated for $D>2$ dimensions. The partition
function is now given by (reintroducing the mass term)
\bea
&& \int_{\bar\phi (0,\vec{x})=\bar\phi (\b ,\vec{x})}
\prod_{\o ,\vec{x}}^{}{d\bar\phi  \bar g^{1\over 4}
(\o ,\vec{x})}e^{-\int_0^{\b}d\o d^{D-1}x\sqrt{\bar g}
[\bar{g^{\mu\nu}}\pa_{\mu}\bar\phi\pa_{\nu}\bar\phi -{1\over
4}(D-2)^2\bar\phi^2+m^2R^2\bar\phi^2 ]}\nn
&&= det^{-{1\over 2}}[-\bar\bbox-{1\over 4}(D-2)^2+m^2R^2] ,
\label{partrind}
\ena
where $\bar\bbox=\pa^2_{\o}+R^{D-1}\pa_R{R^2\over R^{D-1}}\pa_R
+R^2\pa^2_{\bot}$. There is no purely spatial transformation
of coordinates that will bring this into the form of
a partition function for a gas of free bosons at the inverse
temperature $\b$. The second term in (\ref{partrind}) arises from the
conformal coupling due to the fact that the curvature in the
optical space is non-zero. Thus all the terms in the asymptotic
expansion in (\ref{freeenergy}) will contribute.
However, in the high-temperature approximation we will have
(\ref{highfree}) with
\EQ
V_{D-1}=V_{D-2}\int_{\e}^{\infty} {dR\over R^{D-1}}={V_{D-2}\over
(D-2)\e^{D-2}}.
\EN
The last integral is divergent at the origin of optical space and in
the high-temperature approximation; what we have is exactly the free
energy of a gas of massless bosons in a spatial box endowed with
the optical metric, i.e. the optical volume of the box is
${V_{D-2}\over\e^{D-2}}$. This expression for the free energy agrees
with the first calculation done by Susskind and Uglum \cite{suss}.
However we have ignored the terms coming from the (non-zero) optical
space curvature ($B_k, k\ne 0)$ by taking the high-temperature
approximation. These correspond to the contributions also neglected
in the WKB approximation to calculate eigenvalues and in replacing
the sum over modes by an integral in \cite{suss}. Alternatively
they must correspond, in the functional integral calculation of
\cite{suss} in the original metric to finite non-zero contributions
coming from the terms which are of higher order in
curvatures.\footnote{There is a formal argument given in \cite{suss}
to the effect that these terms must vanish. However it involves
delicate issues arising from having products of conical
singularities (delta functions). Indeed it has been argued in
\cite{frolov} on somewhat different grounds
than the above, that these terms must contribute to the entropy.}

\sect{Black Hole Backgrounds}
\indent
The four-dimensional Schwarzschild black hole has the metric (setting
$G_N=1$)
\EQ
ds^2 =- (1-{2M\over r})dt^2 + {dr^2\over (1-{2M\over r})}+r^2d\O_2.
\label{schwarz}
\EN
The corresponding optical volume is
\bea
V_3^{Sch}&=&4\pi\int^R_{2M+\e}{r^2\over  (1-{2M\over
r})^2}\nn &=&4\pi ({R^3\over 3 }+2MR^2
+12M^2R+32M^3\ln{R-2M\over\e}\nn
&+&{16M^4\over\e}-{104M^3\over 3}+O(R^{-1})+O(\e ) ) .
\label{schvol}
\ena
By plugging this into (\ref{highfree}) and (\ref{highfermi}),
we immediately get the (quantum corrections to) the free energy and
hence the entropy of the black hole in the high-temperature
approximation. Here we see the divergence first observed
by~\cite{thooft}. Although it appears linear in terms of the
coordinate cutoff $\e$, it is quadratic in terms of the proper
distance cutoff $\d=\sqrt{2M\e}$ in the Schwarzschild geometry.
We also see another logarithmic divergence. These additional
divergences were first  discovered by working with the functional
integral in the original metric (\ref{partition}). However in that
case the calculation is much more complicated~\cite{solod}.

Next let us consider the Reissner-Nordstr\"om charged black hole.
This example is interesting because it has an extreme limit
when the mass becomes equal to the charge. The metric is
\EQ
ds^2=-(1-{2M\over r}+{Q^2\over r^2})dt^2+(1-{2M\over r}+{Q^2\over
r^2})^{-1}dr^2+r^2d\O_2.
\label{rn}
\EN
This black hole has an ADM mass $M$ and an electric charge $Q$.
The metric has outer and inner horizons at
\EQ
r_{\pm}=M\pm (M^2-Q^2)^{1\over 2}.
\label{hori}
\EN
In order to avoid a naked singularity we must have $M\ge Q$. The
 Hawking temperature of this hole is given by
$T ={(r_+-r_-)\over 4\pi r_+^2}$ (which goes to zero as
$M\to Q$) and the entropy is again given by
the quarter the area of the horizon $S={1\over 4}4\pi r_+^2$ as in
the Schwarzschild case.
In the limit $M\to Q$, the two horizons become degenerate
and the metric of this extreme hole is
\EQ
ds^2=-(1-{M\over r})^2dt^2+{dr^2\over (1-{M\over r})^2}+r^2 d\O_2.
\label{extreme}
\EN
Although the limiting temperature of the Reissner-Nordstr\"om
black hole in the
extreme limit is zero and its entropy is $\pi M^2$, from purely
geometrical considerations in the above metric, it seems that the
temperature is arbitrary and that its entropy is zero~\cite{hor}
even though the area of the horizon is non-zero.
This seems to be rather puzzling from the thermodynamic point of
view.\footnote{We wish to thank L. Susskind for pointing this out.}
We will see below that the calculation of the contribution of the
scalar fields to the entropy sheds some light on this issue.

{}For the non-degenerate case the optical volume is
\bea
V^{rn}_3&=&4\pi\int_{r_++\e}^R {r^6dr\over (r-r_+)^2(r-r_-)^2}
=4\pi [{R^3\over 3} +2MR^2+(3r_+^2+4r_+r_-+3r_-^2)R \nn
&+&{r_+^6\over (r_+-r_-)^2\e} +{r_-^6\over (r_+-r_-)^3}
+{2r_+^5 (2r_+-3r_-)\over (r_+-r_-)^3}\ln{R-r_+ \over \e}\nn
&+&{2r_-^5(3r_+-2r_-)\over (r_+-r_-)^3}\ln {R-r_-\over r_+-r_-}
-r_+({13\over 3}r_+^2+5r_+r_-+3r_-^2)\nn
&+&O(R^{-1})+O(\e)].
\label{rnvol}
\ena
Substituting this in (\ref{highfree}) or (\ref{highfermi}), we get
the expressions for the (quantum corrections to the) free energy and
hence also the entropy in this space. The leading divergence is again
linear (or quadratic in the proper cutoff) and there is an additional
logarithmic divergence. However, we also see the appearance of
inverse powers of the difference in the two horizon radii. This
clearly implies that the extreme limit is very singular. Indeed this
is confirmed by a direct calculation of free energy and entropy for
the extreme black hole. From (\ref{extreme}), we have for
the optical volume
\bea
V^{ext}_3&=&\int_{M+\e}^R {r^6dr\over (r-M)^4}
=4\pi [{R^3\over 3}+2MR^2+10M^2R+{M^6\over 3\e^3}+{3M^5\over\e^2}\nn
&&+{15M^4\over\e}+20M^3\ln{R-M\over\e}-{37\over 3}M^3
+O(R^{-1})+O(\e)].
\label{extvol}
\ena
Here we see the appearance of cubic and quadratic divergences.
Clearly the thermodynamics of the extreme limit is not well-defined
since, although the linear and logarithmic divergences
may be absorbed into the renormalization of $G_N$ \cite{suss} and the
coefficients of higher powers of curvature in the expansion of
the effective action, this will not be the case
for these higher order divergences. We suggest therefore that the
thermodynamics of the extreme limit of the Reissner-Nordstr\"om
black hole is not
well-defined. In fact it would seem that in any discussion of thermal
properties the extreme black hole (which should be able to absorb a
quantum and become non-extreme) must necessarily be treated as the
limiting case of the non-extreme Reissner-Nordstr\"om black
hole.\footnote{It has been claimed \cite{demers} that with Pauli
Villars regulators the entropy of the extreme black hole is no more
singular than that of the non-extreme one. However the authors go on
to point out that the temperature still needs to be taken to be zero
which is in agreement with  our conclusion above.}

Finally we discuss dilaton black holes \cite{gmg}. The metric is
given in this case by
\EQ
ds^2=- (1-{2M\over r})dt^2 + {dr^2\over (1-{2M\over r})}+r(r-a)d\O_2,
\label{dil}
\EN
where $a$ is a constant. The corresponding optical volume is
\bea
V_3&=&4\pi\int^R_{2M+\e}{r(r-a)\over (1-{2M\over r})^2}dr \nn
&=&4\pi ({R^3\over 3 }+(2M-{a\over 2})R^2 +4M(3M-a)R
+{8M^3(2M-a)\over\e}\nn
&+&4M^2(8M-3a)\ln{R-2M\over\e}-M^2({104M\over 3}-10a)
+O(R^{-1})+O(\e)).
\ena
As in the Schwarzschild case here too there is a linear as well as a
logarithmic divergence and again one may argue following \cite{suss}
that the former can be absorbed in a renormalization of $G_N$.
In the extreme limit ($M\to {a\over 2}$), the ``classical"
entropy ($S_{cl}={A\over 4}=2\pi M(2M-a)$~\cite{preskill}) vanishes
and so does the linear divergence. However the logarithmic divergence
remains.

\sect{Conclusions}
\indent
In this paper we have shown that the thermal partition function
is directly related to the functional integral in the so-called
optical metric, and that this differs from the determinant of the
Laplacian in the original metric by a Liouville-type term. As a
consequence we find that the conical singularity method
calculates correctly only  the divergent parts of the entropy. It
should be pointed out that the thermodynamic entropy that we have
discussed is not related in any direct way to the entropy of
entanglement (quantum entropy) that needs to be considered in order
to discuss the issue of whether quantum mechanical information is
lost in the process of black hole evaporation. This effect (or its
absence) is tied up with deep issues in quantum gravity such as the
problem of time and probably also the interpretation of quantum
mechanics itself. What we have studied is the standard
thermostatistical question of counting the number of accessible
micro-states for a macroscopically specified system, in this case
the mass (and charge for the Reissner-Nordstr\"om case) of a black
hole. We have found the quadratic divergence in this entropy that
was found previously but in addition we found a logarithmic
divergence which cannot be interpreted as a renormalization of
Newton's constant. One may of course  subtract this by
introducing a $R^2$ counterterm into the original gravity action
\cite{solod,furs}. Similar results are valid for both
non-extreme charged black holes and for dilatonic black holes.
However in the extreme charged black hole case we have observed
that although the limiting case seems to have no particular problems
the thermodynamics of the limit itself (considered in isolation i.e.
not as the limit of the non-extreme case) does not have well-defined
thermal properties. This is already evident at the classical level
and is exacerbated at the quantum level.

Although it is not possible to understand the issue of quantum
information loss without resolving the deep issues we mentioned
above, one may still hope to understand the operation of the second
law of thermodynamics in this system. Unfortunately a full discussion
of this would involve a solution of a time evolving system and that
is beyond our scope at present. One can however discuss the stability
of the equilibrium between the bath of radiation and the black hole
following~\cite{gp} and other authors.\footnote{For a nice
discussion of these issues and for references to earlier
literature, see~\cite{memb}.} In that work the black hole entropy was
taken to be the classical one and the entropy of the radiation was
taken to be the standard thermodynamic formula. In this work we have
obtained from one calculation both the black hole entropy with
quantum corrections as well as the bulk entropy of the radiation.
After absorbing the divergences by introducing counterterms into
the classical gravitational action (as well as a
$R^2$ term) the same qualitative features will be obtained
(although naturally there will be quantitative differences). Namely
the black hole will be in stable equilibrium with the radiation bath
provided the volume (determined by our infrared cutoff $R$ in the
previous section) is less than a critical value. The calculation
using the conical singularity method, on the other hand, will not
easily yield the (bulk) contribution of the radiation to the
entropy (i.e. a $R^3T^3$ piece) and thus cannot by itself (i.e.
ignoring the Liouville-type term and possible non-local terms)
satisfy the second law.
\sect{Acknowledgments}
\indent
SdeA would like to acknowledge the award of a Japan Society for the
Promotion of Science  fellowship, and the hospitality of Profs.
K. Higashijima and E. Takasugi at Osaka University where this
investigation was begun. He would also like to thank Gary Gibbons and
Lenny
Susskind for discussions on extremal black holes and entropy, and
Doug
MacIntire for discussions on this work. This work is partially
supported by the
Department of Energy contract No. DE-FG02-91-ER-40672.

\end{document}